\documentclass[10pt,a4paper]{amsart}
\usepackage{amsmath,amsthm,amssymb,amsfonts}

\newtheorem*{icet}{$I$CET}
\newtheorem*{gcp}{GCP}
\newtheorem*{mucet}{$\mu$CET}
\newtheorem*{cwlln}{CWLLN}
\newtheorem*{mucwlln}{$\mu$CWLLN}
\newtheorem*{egcp}{EGCP}
\newtheorem*{bcp}{BCP}
\newtheorem*{jcet}{$J$CET}
\newtheorem*{mm}{MaxProb/MaxEnt}
\newtheorem*{mugcp}{$\mu$GCP}
\newtheorem*{lem}{Lemma}

 \theoremstyle{remark}
 \newtheorem*{notes}{Notes}

\begin{document}

\title{Gibbs conditioning extended, \\ Boltzmann conditioning introduced}
\author{M. Grendar}
\address{ Institute of Mathematics and CS of Mathematical Institute of Slovak Academy of Sciences (SAS) and of Matej Bel University,
 Severna 5,
 974 01 Banska Bystrica, Slovakia and Institute of Measurement Science of SAS, Dubravska cesta 9, 841 04 Bratislava, Slovakia}
 \email{marian.grendar@savba.sk }
 \dedicatory{To Mar, in memoriam}
\thanks{This work was supported by VEGA 1/0264/03 grant. Valuable discussions with Brian R. La Cour, Alberto Solana-Ortega,
Ondrej \v Such and Viktor Witkovsk\'y  are gratefully acknowledged.}
 \date{September 17, 2004 revision of August 12, 2004 revision of July 10, 2004 version}
% \subjclass{ }
 \keywords{multiple $I$-projections, Conditioned Weak Law of Large Numbers, Gibbs Conditioning Principle,
 $\mu$-projection, $J$-projection, $\gamma$-projection, Maximum Probability method, MaxProb/MaxEnt convergence}
 \begin{abstract}
Conditional Equi-concentration of Types on $I$-projections ($I$CET)
and Extended Gibbs Conditioning Principle (EGCP) provide an
extension of Conditioned Weak Law of Large Numbers and of Gibbs
Conditioning Principle to the case of non-unique Relative Entropy
Maximizing (REM) distribution (aka $I$-projection). $I$CET and EGCP
give a probabilistic justification to REM under rather general
conditions. $\mu$-projection variants of the results are introduced.
They provide a probabilistic justification to Maximum Probability
(MaxProb) method. 'REM/MaxEnt or MaxProb?' question is discussed
briefly. Jeffreys Conditioning Principle is mentioned.
\end{abstract}

\maketitle

\section{Introduction}

Relative entropy maximization  (REM/MaxEnt) is usually performed
under moment consistency constraints. The constraints define a
feasible set of probability distributions which is convex, closed
and hence the relative entropy maximizing distribution (aka
$I$-projection) is unique. For such  sets Conditioned Weak Law of
Large Numbers (CWLLN) is established and provides a probabilistic
justification of REM/MaxEnt. Gibbs conditioning principle (GCP) - a
stronger version of CWLLN - which is as well established for such
sets, gives a further insight into the 'phenomenon' of conditional
concentration of empirical measure on $I$-projections.

This work strives to develop  extensions of CWLLN  and GCP to the
case of non-unique $I$-projection\footnote{For a motivation see
\cite{gg_nonlin}, \cite{LS}. For an exploratory work in this
direction see \cite{gg_aei}.}. Proposed Conditional
Equi-concentration of Types on $I$-projections ($I$CET) which
extends CWLLN says, informally, that types (i.e., empirical
distributions) conditionally concentrate on each of proper
$I$-projections in equal measure. Extended Gibbs conditioning
principle (EGCP) states, that in the case of multiple proper
$I$-projections, probability of an outcome is given by equal-weight
mixture of proper $I$-projection probabilities of the outcome.

A generalization (cf. \cite{gg_asy}) of a result on convergence of
maximum/supremum  probability types ($\mu$-projections) to
$I$-projections  (cf. \cite{gg_what}, Thm 1) %, aka MaxProb/MaxEnt Thm)
directly permits to state either the well-established CWLLN, GCP or
their extensions equivalently in terms of $\mu$-projections. The
$\mu$-projection variants of the probabilistic laws allows  for a
deeper reading than their $I$-projection counterparts - since the
$\mu$-laws express the asymptotic conditional behavior of types in
terms of the asymptotically most probable types. They provide
probabilistic justification to Maximum Probability (MaxProb) method.

Though $\mu$-projections and $I$-projections are asymptotically
identical, in the case of finite samples, they are in general
different.

\section{Terminology and notation}

Let $\{X\}_{l=1}^n$ be a sequence of independently and identically
distributed random variables with a common law (measure)  on a
measurable space. Let the measure be concentrated on $m$ atoms from
a set $\mathrm{X} \triangleq \{x_1, x_2, \dots, x_m\}$ called
support or alphabet. Hereafter $\mathrm{X}$ will be assumed finite.
An element of $\mathrm{X}$ will be called outcome or letter. Let
$q_i$ denote the probability (measure) of $i$-th element of $\mathrm
X$; $q$ will be called source or generator. Let
$\mathrm{P}(\mathrm{X})$ be a set of all probability mass functions
(pmf's) on $\mathrm X$.

A type  (also called $n$-type, empirical measure, frequency
distribution or occurrence vector) induced by a sequence
$\{X\}_{l=1}^n$ is the pmf $\nu^n \in \mathrm{P}(\mathrm{X})$ whose
$i$-th element $\nu_i^n$ is defined as: $\nu_i^n \triangleq n_i/n$
where $n_i \triangleq \sum_{l=1}^n I(X_l = x_i)$, and $I(\cdot)$ is
the characteristic function.  Multiplicity $\Gamma(\nu^n)$ of type
$\nu^n$ is: $\Gamma(\nu^n) \triangleq n!/ \prod_{i=1}^m n_i!$.

Let $\mathrm{\Pi} \subseteq \mathrm{P}(\mathrm{X})$.  Let
$\mathrm{P}_n$ denote a subset of $\mathrm{P}(\mathrm{X})$ which
consists of all $n$-types. Let $\mathrm{\Pi}_n = \mathrm\Pi \cap
\mathrm{P}_n$.

$I$-projection $\hat p$ of $q$ on $\mathrm\Pi$ is  $\hat p
\triangleq \arg \, \inf_{p \in \mathrm\Pi} I(p || q)$, where $I(p ||
q) \triangleq \sum_{\mathrm{X}} p_i \log\frac{p_i}{q_i}$ is
Kullback-Leibler distance, information divergence or minus relative
entropy.

$\pi(\nu^n \in \mathrm{A}| \nu^n \in \mathrm{B}; q \mapsto \nu^n)$
will denote the conditional probability that if a type drawn from $q
\in \mathrm{P}({\mathrm X})$ belongs to $\mathrm{B} \subseteq
\mathrm{\Pi}$ then it belongs to $\mathrm{A} \subseteq
\mathrm{\Pi}$.

\section{CWLLN, Gibbs conditioning}

Conditioned Weak Law of Large Numbers (cf. \cite{Bartfai},
\cite{Jaynes}, \cite{Vasicek}, \cite{CC}, \cite{Csi}, \cite{Ellis},
\cite{H}) in its standard form (cf. \cite{CT}) reads:

\begin{cwlln}
 Let $\mathrm{X}$ be a finite set. Let $\mathrm{\Pi}$ be
closed, convex set which does not contain $q$. Let $n \rightarrow
\infty$. Then for $\epsilon
> 0$
$$
\lim_{n \rightarrow \infty} \pi(|\nu^n_i - \hat{p}_i| < \epsilon |
\nu^n \in \mathrm{\Pi}; q \mapsto \nu^n) = 1 \text{\ \ for\ } i = 1,
2, \dots, m.
$$
\end{cwlln}

%An information-theoretic  proof (see \cite{CT}) of CWLLN utilizes
%so-called Py\-tha\-go\-re\-an theorem (cf. \cite{Chentsov}),
%Pin\-sker inequality and standard inequalities for factorial.
%Alternatively, CWLLN can be obtained as a consequence of Sanov
%Theorem.

CWLLN says that if types  are confined to a closed, convex set
$\mathrm\Pi$ then they asymptotically conditionally concentrate on
the $I$-projection $\hat{p}$ of the source of types $q$ on the set
$\mathrm\Pi$ (i.e., informally,  on the probability distribution
from $\mathrm\Pi$ which has the highest value of the relative
entropy with respect to the source $q$).

Gibbs conditioning principle (GCP) says, very informally, that if
the source $q$ is confined to produce sequences which lead to types
in a convex, clsoed set $\mathrm\Pi$ then elements of any such
sequence (of fixed length $t$) behave asymptotically conditionally
as if they were drawn identically and independently from the
$I$-projection of $q$ on $\mathrm\Pi$ - provided that the last is
unique (among other things).

\begin{gcp}
 Let $\mathrm{X}$ be a finite set. Let $\mathrm{\Pi}$ be closed,
convex set which does not contain $q$. Let $n \rightarrow \infty$.
Then for a fixed $t$
$$
\lim_{n \rightarrow \infty} \pi(X_1 =x_1, \dots, X_t = x_t | \nu^n
\in \mathrm{\Pi}; q \mapsto \nu^n) = \prod_{l=1}^t \hat{p}_{x_l}.
$$
\end{gcp}

GCP was developed at \cite{Csi} under the name of conditional
quasi-independence of outcomes. Later on, it was brought into more
abstract form in large deviations literature, where it also obtained
the GCP name (cf. \cite{DZ}, \cite{LN}). A simple proof of GCP can
be found at \cite{CsiMT}. GCP is proven also for continuous alphabet
(cf. \cite{GOR},  \cite{CsiMT}, \cite{DZ}).

\section{The case of several $I$-projections}

What happens when $\mathrm\Pi$ admits multiple $I$-projections? Do
the conditional concentration of types happens on them? If yes, do
types concentrate on each of them? If yes, what is the proportion?
How does GCP extend to the case of multiple $I$-projections?

\subsection{Conditional Equi-concentration of Types on
$I$-projections}

Let $d(a,b)$ $\triangleq \sum_{i = 1}^m |a_i - b_i|$ be the total
variation metric (or any other equivalent metric) on the set of
probability distributions $\mathrm{P}(\mathrm{X})$. Let $B(a,
\epsilon)$ denote an $\epsilon$-ball - defined by the metric $d$ -
which is centered at $a \in \mathrm{P}(\mathrm{X})$.

An $I$-projection $\hat{p}$ of $q$ on $\mathrm\Pi$ will be called
proper if  $\hat{p}$ is not an isolated point of $\mathrm\Pi$.

\smallskip

\begin{icet}
 Let $\mathrm{X}$ be a finite set. Let $\mathrm\Pi$ be such that it admits $\mathrm k$ proper
$I$-projections $\hat{p}^1, \hat{p}^2, \dots, \hat{p}^\mathrm{k}$ of
$q$. Let $\epsilon > 0$ be such that for $j = 1, 2, \dots,
\mathrm{k}$ $\hat{p}^j$ is the only proper $I$-projection of $q$ on
$\mathrm\Pi$ in the ball $B(\hat{p}^j, \epsilon)$. Let $n
\rightarrow \infty$. Then
\begin{equation}
\pi(\nu^n  \in B(\epsilon, \hat{p}^j) | \nu^n \in \mathrm{\Pi}; q
\mapsto \nu^n) = 1/\mathrm{k} \text{\ \ for\ } j = 1, 2, \dots,
\mathrm{k}.
\end{equation}
\end{icet}

 $I$CET\footnote{See Appendix for a proof of $I$CET and EGCP.}  states that if a set $\mathrm\Pi$ admits several
$I$-projections then the conditional measure is spread among the
proper $I$-projections equally. In less formal words:  if a random
generator (i.e., $q$) is confined to produce types in  $\mathrm\Pi$
then, as $n$ gets large, the generator 'hides itself' equally likely
behind any of its proper $I$-projections on $\mathrm\Pi$. Yet in
other (statistical physics) words: each of the equilibrium points
(i.e., proper $I$-projections) is asymptotically conditionally
equally probable. The conditional equi-concentration of types
'phenomenon' resembles Thermodynamic coexistence of phases (e.g.,
triple point of water, vapor and ice).

\begin{notes}
 1) On an $I$-projection $\hat{p}$ which is not rational and at the same time it is an isolated point no conditional
concentration of types happens. However, if the  set $\mathrm\Pi$ is
such that an $I$-projection $\hat{p}$ of $q$ on it is rational and
at the same time it is an isolated point, then types can concentrate
on it. 2) Since $\mathrm{X}$ is finite, $\mathrm k$ is finite.
\end{notes}

 Weak Law of Large Numbers is special - unconditional - case of
CWLLN. CWLLN itself is just a special - unique proper $I$-projection
- case of $I$CET.

Two illustrative examples of the Conditional Equi-concentration of
Types on $I$-projections ($I$CET) can be found at the exploratory
study \cite{gg_aei}. There also Asymptotic Equiprobability of
$I$-projections - a precursor to $I$CET - was formulated.

\subsection{Extended Gibbs conditioning principle}

\begin{egcp}
 Let $\mathrm{X}$ be  a finite set.
 Let $\mathrm\Pi$ be such that it admits $\mathrm k$ proper $I$-projections $\hat{p}^1,
\hat{p}^2, \dots, \hat{p}^\mathrm{k}$ of $q$ on $\mathrm\Pi$. Then
for a fixed $t$:
\begin{equation}
\lim_{n \rightarrow \infty} \pi(X_1 = x_1, \dots, X_t = x_t | \nu^n
\in \mathrm\Pi; q \mapsto \nu^n) = 1/\mathrm{k}
\sum_{j=1}^{\mathrm{k}} \prod_{l=1}^t \hat{p}_{x_l}^j.
\end{equation}
\end{egcp}

EGCP, for $t=1$, says that the conditional probability of  a letter
is asymptotically given by the equal-weight mixture of proper
$I$-projection probabilities of the letter. For a general sequence,
EGCP states that the conditional probability of a sequence is
asymptotically equal to the mixture of joint probability
distributions. Each of the $\mathrm{k}$ joint distributions is such
as if the sequence was iid distributed according to a proper
$I$-projection.

\section{$\mu$-projections, Maximum Probability method}

$\mu$-projection  $\hat{\nu}^n$ of $q$ on $\mathrm{\Pi}_n \neq
\emptyset$ is defined as: $\hat{\nu}^n \triangleq \arg \,
\sup_{\nu^n \in \mathrm{\Pi}_n} \pi(\nu^n; q)$, where $\pi(\nu^n; q)
\triangleq \Gamma(\nu^n) \prod (q_i)^{n \nu^n_i}$, (cf.
\cite{gg_asy}). Alternatively, the $\mu$-projection can be defined
as $\hat{\nu}^n \triangleq \arg \, \sup_{\nu^n \in \mathrm{\Pi}_n}
\pi(\nu^n | \nu^n \in \mathrm{\Pi}_n; q)$, where $\pi(\nu^n| \nu^n
\in \mathrm{\Pi}_n; q)$ denotes the conditional probability that if
an $n$-type belongs to $\mathrm{\Pi}_n$ then it is just the type
$\nu^n$. Yet another equivalent definition - a bayesian one - of
$\mu$-projection can be adapted from \cite{gg_bayes}.

Concept of $\mu$-projection is associated with the Maximum
Probability method (cf. \cite{gg_what}).

\subsection{Asymptotic identity of $\mu$-projections and $I$-projections}

At (\cite{gg_what}, Thm 1 and its Corollary, aka MaxProb/MaxEnt Thm)
it was shown that maximum probability type converges to
$I$-projection; provided that $\mathrm{\Pi}$ is defined by a
differentiable constraints. A more general result which states
asymptotic identity of $\mu$-projections and $I$-projections  was
presented at \cite{gg_asy}. It will be recalled here.

\begin{mm}
{\rm \cite{gg_asy}}  Let $\mathrm{X}$ be a finite set. Let
$\mathrm{M}_n$ be set of all $\mu$-projections of $q$ on
$\mathrm{\Pi}_n$. Let $\mathrm{I}$ be set of all $I$-projections of
$q$ on $\mathrm{\Pi}$. For $n \rightarrow \infty$, $\mathrm{M}_n =
\mathrm{I}$.
\end{mm}

Since $\pi(\nu^n; q)$ is defined for $\nu^n \in \mathrm{Q}^m$,
$\mu$-projection can be defined only for $\mathrm{\Pi}_n$ when $n$
is finite. The Theorem permits to define a $\mu$-projection
$\hat{\nu}$ also on $\mathrm{\Pi}$: $\hat{\nu} \triangleq \arg
\sup_{r \in \mathrm{\Pi}} - \sum_{i=1}^m$ $r_i \log\frac{r_i}{q_i}$.
Thus $\mu$-projections  and $I$-projections on $\mathrm{\Pi}$  are
undistinguishable.

It is worth highlighting that for a finite $n$, $\mu$-projections
and $I$-projections of $q$ on  $\mathrm{\Pi}_n$ are in general
different. This explains why $\mu$-form of the probabilistic laws
deserves to be stated separately of the $I$-form; though formally
they are undistinguishable. Thus,  MaxProb/MaxEnt Thm (in its new
and to a smaller extent also in its old version) permits directly to
state $\mu$-projection variants of CWLLN, GCP, $I$CET and EGCP:
$\mu$CWLLN, $\mu$GCP, $\mu$CET and Boltzmann Conditioning Principle
(BCP).

\subsection{$\mu$-form of CWLLN and GCP}

\begin{mucwlln}
 Let $\mathrm{X}$ be a finite set. Let $\mathrm{\Pi}$ be closed,
convex set which does not contain $q$. Let $n \rightarrow \infty$.
Then for $\epsilon
> 0$
$$
\lim_{n \rightarrow \infty} \pi(|\nu^n_i - \hat{\nu}_i| < \epsilon |
\nu^n \in \mathrm{\Pi}; q \mapsto \nu^n) = 1 \text{\ \ for\ } i = 1,
2, \dots, m.
$$
\end{mucwlln}

Core of $\mu$CWLLN can be loosely expressed as: {\it types, when
confined to a convex, closed set} $\mathrm{\Pi}$, {\it conditionally
concentrate on the asymptotically most probable type} $\hat{\nu}$.
It is worth a comparison with the reading of  the $I$-projection
variant of CWLLN (see Sect.~3).

Similarly, to the GCP its  $\mu$-variant exists:

\begin{mugcp}
 Let $\mathrm{X}$ be a finite set. Let $\mathrm{\Pi}$ be
closed, convex set which does not contain $q$. Let $n \rightarrow
\infty$. Then for a fixed $t$
$$
\lim_{n \rightarrow \infty} \pi(X_1 =x_1, \dots, X_t = x_t | \nu^n
\in \mathrm{\Pi}; q \mapsto \nu^n) = \prod_{l=1}^t \hat{\nu}_{x_l}.
$$
\end{mugcp}

\subsection{Conditional Equi-concentration of Types on
$\mu$-projections}

A $\mu$-pro\-ject\-ion $\hat{\nu}$ of $q$ on $\mathrm\Pi$ will be
called proper if  $\hat{\nu}$ is not an isolated point of
$\mathrm\Pi$.

\begin{mucet}
 Let $\mathrm{X}$ be a finite set. Let there be $\mathrm k$
proper $\mu$-projections $\hat{\nu}^1, \hat{\nu}^2, \dots,
\hat{\nu}^\mathrm{k}$ of $q$ on $\mathrm\Pi$. Let $\epsilon > 0$ be
such that for $j = 1, 2, \dots, \mathrm{k}$ $\hat{\nu}^j$ is the
only proper $\mu$-projection of $q$ on $\mathrm\Pi$ in the ball
$B(\hat{\nu}^j, \epsilon)$. Let $n \rightarrow \infty$. Then
\begin{equation}
\pi(\nu^n  \in B(\epsilon, \hat{\nu}^j) | \nu^n \in \mathrm{\Pi}; q
\mapsto \nu^n) = 1/\mathrm{k} \text{\ \ for\ } j = 1, 2, \dots,
\mathrm{k}.
\end{equation}
\end{mucet}

\subsection{Boltzmann conditioning principle}

\begin{bcp}
 Let $\mathrm{X}$ be a finite set. Let there be $\mathrm k$ proper $\mu$-projections
$\hat{\nu}^1, \hat{\nu}^2, \dots, \hat{\nu}^\mathrm{k}$ of $q$ on
$\mathrm\Pi$. Then for a fixed $t$:
\begin{equation}
\lim_{n \rightarrow \infty} \pi(X_1 = x_1, \dots, X_t = x_t | \nu^n
\in \mathrm\Pi; q \mapsto \nu^n) =  1/\mathrm{k}
\sum_{j=1}^{\mathrm{k}} \prod_{l=1}^t \hat{\nu}_{x_l}^j.
\end{equation}
\end{bcp}

\subsection{MaxEnt or MaxProb?}

$\mu$-projections and $I$-projections are asymptotically
indistinguishable (recall MaxProb/MaxEnt Thm, Sect. 5.1). In plain
words: for $n \rightarrow \infty$ REM/MaxEnt selects the same
distribution(s) as MaxProb (in its more general form which instead
of the maximum probable types selects supremum-probable
$\mu$-projections). This result (in the older form, \cite{gg_what})
was at \cite{gg_what} {\it interpreted} as saying that REM/MaxEnt
can be viewed as an asymptotic instance of the simple and
self-evident Maximum Probability method.

Alternatively, \cite{Alberto} suggests to view REM/MaxEnt as a
separate method and hence to read the MaxProb/MaxEnt Thm as claiming
that REM/MaxEnt asymptotically coincides with MaxProb. If one adopts
this interesting and legitimate view then it is necessary to face
the fact that if $n$ is finite, the two methods in general differ.

\section{Jeffreys conditioning mentioned}

In\-stead of Summ\-ary (which is al\-rea\-dy pre\-sent\-ed at Sect.
1), Con\-di\-tio\-nal Equi-con\-cen\-tra\-ti\-on of types on
$J$-projections ($J$CET) and Jeffreys conditioning
principle\footnote{It should not be confused with Jeffrey principle
of updating subjective probability.} (JCP) will be mentioned, in
passing.

$\gamma$-projection $\tilde{\nu}^n$ of $q \in \mathrm{Q}^m$ on
$\mathrm{\Pi}_n$ is: $\tilde{\nu}^n \triangleq \arg \sup_{\nu^n \in
\mathrm{\Pi}_n} \pi(\nu^n; q) \, \pi(q; \nu^n)$. $J$-projection (or
Jeffreys projection) $\tilde{p}$ of $q \in \mathrm{Q}^m$ on
$\mathrm\Pi$ is $\tilde{p} \triangleq \arg \inf_{p \in \mathrm\Pi}
\sum_{i=1}^m p_i \log \frac{p_i}{q_i} + q_i \log \frac{q_i}{p_i}$.

Let $q \in \mathrm{Q}^m$. $\pi(\nu^n \in \mathrm{A}| \nu^n \in
\mathrm{B}; (q \mapsto \nu^n) \wedge (\nu^n \mapsto q))$ will denote
the conditional probability that if a type - which was drawn from $q
\in \mathrm{P}({\mathrm X})$ and was at the same time used as a
source of the type $q$ - belongs to $\mathrm{B} \subseteq
\mathrm{\Pi}$ then it belongs to $\mathrm{A} \subseteq
\mathrm{\Pi}$. A $J$-projection $\tilde{p}$ of $q$ on $\mathrm\Pi$
will be called proper if  it is not isolated point of $\mathrm\Pi$.

\begin{jcet}
 Let $\mathrm{X}$ be a finite set.
 Let $q \in \mathrm{Q}^m$. Let there be $\mathrm k$ proper
$J$-projections $\tilde{p}^1, \tilde{p}^2, \dots,
\tilde{p}^\mathrm{k}$ of $q$ on $\mathrm\Pi$. Let $\epsilon > 0$ be
such that for $j = 1, 2, \dots, \mathrm{k}$ $\tilde{p}^j$ is the
only proper $J$-projection of $q$ on $\mathrm\Pi$ in the ball
$B(\tilde{p}^j, \epsilon)$.  Let $n_0$ be denominator of the
smallest common divisor of $q_1, q_2, \dots, q_m$. Let $n = u n_0$,
$u \in \mathrm{N}$. Let $u \rightarrow \infty$ . Then
\begin{equation}
\pi(\nu^n  \in B(\epsilon, \tilde{p}^j) | \nu^n \in \mathrm{\Pi}; (q
\mapsto \nu^n) \wedge (\nu^n \mapsto q)) = 1/\mathrm{k} \text{\ \
for\ } j = 1, 2, \dots, \mathrm{k}.
\end{equation}
\end{jcet}

In words, types which were 'emitted' from $q$ and were at the same
time used as a source of $q$-types, conditionally equi-concentrate
on $J$-projections of $q$ on  $\mathrm\Pi$.

$J$-projections and $\gamma$-projections asymptotically coincide
(cf. \cite{gg_asy}, and \cite{gg_ij} for an example). Hence, a
$\gamma$-projection alternative of $J$CET is valid as well. It says
that: types which were 'emitted' from $q$ and were at the same time
used as a source of $q$-types, conditionally equi-concentrate on
those of them which have the highest/supremal value of $\pi(\nu^n;
q)\, \pi(nq; \nu^n)$. -- Similarly, JCP can be considered in its
$J$- or $\gamma$-form.

$\mu$-projection is based on the probability $\pi(\nu^n; q)$; thus
it can be viewed as a $UNI$-projection. $\gamma$-projection is based
on $\pi(\nu^n; q)\, \pi(nq; \nu^n)$, thus it can be viewed as
$AND$-projection. It is possible to consider also an $OR$-projection
defined as $\dot{\nu}^n \triangleq \arg \sup_{\nu^n \in
\mathrm{\Pi}_n} \pi(\nu^n; q) + \pi(nq; \nu^n)$. However, there
seems to be no obvious analytic way how to define its asymptotic
form. Despite that, it is possible to expect that $OR$-type of
CWLLN/CET holds.

The $\mu$-, $\gamma$-, $OR$-projection CET can be summarized by a
(bold) statement: types conditionally equi-concentrate on those
which are asymptotically the most probable.

\medskip

{\bf Acknowledgments} {\ Supported by VEGA 1/0264/03. It is a
pleasure to thank Brian R. La Cour, Alberto Solana-Ortega, Ondrej \v
Such and Viktor Witkovsk\'y for valuable discussions. Lapses are
mine. The author is indebted to the Editors for patience.}

%\appendix
\section{Appendix}

\smallskip

\begin{proof}[A sketch of proof of $I$CET]

\begin{equation}
\pi(\nu^n \in B(\epsilon, \hat{p}^j) | \nu^n \in \mathrm{\Pi}; q
\mapsto \nu^n) \le \frac{ \sum_{\nu^n \in B} \pi(\nu^n;
q)}{\sum_{\nu^n \in \mathrm\Pi} \pi(\nu^n; q)}
\end{equation}

$B_n(\epsilon, \hat{p}^j) \triangleq B(\epsilon, \hat{p}^j) \cap
\mathrm{\Pi}_n$. Let there be $\mathrm{k}_{B,n}$ $I$-projections
$\hat{p}_{B,n}^1, \hat{p}_{B,n}^2, \dots,
\hat{p}_{B,n}^{\mathrm{k}_{B,n}}$ of $q$ on
$\bigcup_{j=1}^\mathrm{k} B_n(\epsilon, \hat{p}^j)$. Let
$\mathrm{k}_{B,n}^j$ denote the number of $I$-projections of $q$ on
$B_n(\epsilon, \hat{p}^j)$. $\hat{p}_{{B}, n}^j$ will stand for any
of such $I$-projections. Denote the set $B_n(\epsilon, \hat{p}_n^j)
\backslash \bigcup_{i=1}^{\mathrm{k}_{B,n}^j} \{\hat{p}_{{B},
n}^i\}$ as $B\backslash{\mathrm{k}_{B,n}^j}$.

Similarly, let there be $\mathrm{k}_{\mathrm{\Pi}, n}$
$I$-projections $\hat{p}_{\mathrm{\Pi}, n}^1, \hat{p}_{\mathrm{\Pi},
n}^2, \dots, \hat{p}_{\mathrm\Pi, n}^{\mathrm{k}_n}$ of $q$ on
$\mathrm{\Pi}_n$. Denote the set $\mathrm{\Pi}_n \backslash
\bigcup_{i=1}^{\mathrm{k}_{\mathrm{\Pi}, n}}
\{\hat{p}_{\mathrm{\Pi}, n}^i\}$ as
$\mathrm{\Pi}\backslash{\mathrm{k}_{\mathrm{\Pi}, n}}$.
The MaxProb/MaxEnt Thm implies that for $n \rightarrow \infty$ the
RHS of (6) can be
 written as:
\begin{equation}
\frac{ \pi(\hat{p}_{B,n}^j; q) \left( \mathrm{k}_{{B}, n}^j +
 \frac{\sum_{\nu^n \in B\backslash{\mathrm{k}_{{B}, n}^j}} \pi(\nu^n; q)}{\pi(\hat{p}_{{B} ,n}^j; q)
}\right)}{\pi(\hat{p}_{\mathrm{\Pi}, n}; q)
\left(\mathrm{k}_{\mathrm{\Pi}, n} + \frac{\sum_{\nu^n \in
\mathrm{\Pi}\backslash{\mathrm{k}_{\mathrm{\Pi}, n}}} \pi(\nu^n;
q)}{\pi(\hat{p}_{\mathrm{\Pi},n}; q) }\right)}
\end{equation}

Recall  a standard inequality:
\begin{lem}
  Let $\nu^n$, $\dot{\nu}^n$ be two types from $\mathrm{\Pi}_n$.
Then
\begin{equation*}
\frac{\pi(\nu^n; q)}{\pi(\dot{\nu}^n; q)} <
\left(\frac{n}{m}\right)^m \prod_{i=1}^m \frac{
(\frac{q_i}{\nu_i^n})^{n \nu_i^n} }{ (\frac{q_i}{\dot{\nu}_i^n})^{n
\dot{\nu}_i^n}  }
\end{equation*}
\end{lem}

The Lemma  implies that the ratio in the nominator of (7) converges
to zero as $n \rightarrow \infty$. The same implication holds for
the ratio in the denominator. $\hat{p}_{B, n}^j$ converges in the
metric to $\hat{p}^j$, hence $\mathrm{k}_{B,n}^j$ converges to $1$
as $n \rightarrow \infty$. Similarly, $\mathrm{k}_{\mathrm{\Pi}, n}$
converges to $\mathrm{k}$ and $\frac{\pi(\hat{p}_{B,n}^j;
q)}{\pi(\hat{p}_{\mathrm{\Pi}, n}; q)}$ converges to $1$ as $n$ goes
to infinity. This taken together implies that the RHS of (6)
converges  to $1/\mathrm{k}$ as $n \rightarrow \infty$. The
inequality (6) thus turns into equality.
\end{proof}

%------------------------------ EGCP
\begin{proof}[A sketch of proof of EGCP]
\begin{equation}
\pi(X_1 = x_1, \dots, X_t = x_t | \nu^n \in \mathrm\Pi; q \mapsto
\nu^n) = \frac{\sum_{\nu^n \in \mathrm\Pi} \pi(X_1 = x_1, \dots, X_t
= x_t, \nu^n)}{\sum_{\nu^n \in \mathrm\Pi} \pi(\nu^n; q)}
\end{equation}

Partition $\mathrm{\Pi}_n$ into
$\mathrm{\Pi}\backslash{\mathrm{k}_{\mathrm{\Pi},n}}$ and the rest,
which will be denoted by $\bigcup \hat{p}_{\mathrm{\Pi},n}$. The
MaxProb/MaxEnt Thm implies that for $n \rightarrow \infty$ the RHS
of (8) can be written as:

\begin{equation}
\frac{\sum_{\nu^n \in \bigcup \hat{p}_{\mathrm{\Pi},n}} \pi(X_1 =
x_1, \dots, X_t = x_t, \nu^n) + \sum_{\nu^n \in
\mathrm\Pi\backslash\mathrm{k}_{\mathrm{\Pi}, n}} \pi(X_1 = x_1,
\dots, X_t = x_t, \nu^n)}{\pi(\hat{p}_{\mathrm{\Pi}, n};
q)(\mathrm{k}_{\mathrm{\Pi}, n}
 + \frac{\sum_{\nu^n \in \mathrm\Pi\backslash\mathrm{k}_{\mathrm{\Pi},
n}} \pi(\nu^n; q)}{\pi(\hat{p}_{\mathrm{\Pi}, n}; q)})}
\end{equation}

By the Lemma, the ratio in the denominator of (9) converges to zero
as $n$ goes to infinity. The second term in the nominator as well
goes to zero as $n \rightarrow \infty$ (to see this, express the
joint probability $\pi(X_1 = x_1, \dots, X_t = x_t, \nu^n)$ as
$\pi(X_1 = x_1, \dots, X_t = x_t|\nu^n)\pi(\nu^n; q)$ and employ the
Lemma). Thus, for $n \rightarrow \infty$ the RHS of (8) becomes
$1/\mathrm{k} \sum_{j=1}^{\mathrm{k}} \pi(X_1 = x_1, \dots, X_t =
x_t|\hat{p}^j)$. Finally, invoke Csisz\'ar's 'urn argument' (cf.
\cite{CsiMT}) to conclude that the asymptotic form of the RHS of (8)
is $1/\mathrm{k} \sum_{j=1}^{\mathrm{k}} \prod_{l=1}^t
\hat{p}_{X_l}^j$.
\end{proof}

%\smallskip

\newpage

\section{Changes wrt the Version 3}

Three major changes: 1) Definition of proper $I$-projection has been
changed. 2) An argument preceding Eq. (7) at the proof of $I$CET
(and similarly Eq. (9) at the proof of EGCP) is now correctly
stated. 3) Abstract was rewritten to better reflect contents of
paper.

This is the definitive form of the work. To appear at the
Proceedings of MaxEnt'04 workshop.

\end{document}